\documentclass[a4paper,showpacs,nofootinbib,twocolumn]{revtex4-1}
\usepackage[utf8]{inputenc}
\usepackage{graphicx,bm,color,amssymb,amsmath,natbib,epsfig,hhline}
\usepackage{slashed,verbatim,ulem}
\usepackage{subfigure}
\usepackage[colorlinks=true,linktocpage=true,linkcolor=blue,citecolor=blue]{hyperref}
\definecolor{mypink}{RGB}{219, 48, 222}
\definecolor{brown}{RGB}{200,150, 10}

\def\be {\begin{equation}}
\def\ee {\end{equation}}
\def\bea {\begin{eqnarray}}
\def\eea {\end{eqnarray}}
\def\bc {\begin{center}}
\def\ec {\end{center}}
\def\nn {\nonumber}

\date{\today}

\begin{document}
\title{{GW170817: constraining the  nuclear matter equation of state from the neutron star tidal deformability}}
\author{Tuhin Malik$^1$}
\email{tuhin.malik@gmail.com}
\author{N. Alam$^{2,3}$}
\author{M. Fortin$^4$}
\author{C. Provid{\^e}ncia$^5$}
\author{B. K. Agrawal$^{2,6}$ \break}
\author{T. K. Jha$^{1}$}
\author{Bharat Kumar$^{7,6}$}
\author{S. K. Patra$^{7,6}$}

\affiliation{$^1$Department of Physics, BITS-Pilani, K.K. Birla Goa Campus, Goa - 403726, India\\
$^2$Saha Institute of Nuclear physics, Kolkata 700064, India\\
$^3$ Theoretical Physics Division, Variable Energy Cyclotron Centre, 1/AF Bidhannagar, Kolkata 700064, India \\
$^4$N. Copernicus Astronomical Center, Polish Academy of Science,
Bartycka,18, 00-716 Warszawa, Poland\\
$^{5}$CFisUC, Department of Physics, University of Coimbra, 3004-516 Coimbra, Portugal\\
$^6$Homi Bhabha National Institute, Anushakti Nagar, Mumbai -
400094, India. \\
$^{7}$Institute of Physics, Bhubaneswar - 751005, India.}
\begin{abstract}
{Constraints set on key parameters of the nuclear matter equation of state (EoS) by the values 
of the tidal deformability, inferred from GW170817, are examined by using a diverse set of relativistic 
and non-relativistic mean field models. These models are consistent with bulk properties of finite 
nuclei as well as with the observed lower bound on the maximum mass of neutron star $\sim 2 ~ {\rm M}_\odot$. 
The tidal deformability shows a strong correlation with specific linear combinations of the isoscalar and isovector
nuclear matter parameters associated with the EoS. Such correlations suggest that a precise value of the tidal deformability
can put {tight} bounds on several EoS parameters, in particular, on the slope of the incompressibility 
and the curvature of the symmetry energy. The tidal deformability obtained from the  GW170817 and its  
UV/optical/infrared counterpart  sets the radius of a canonical $1.4~ {\rm M}_{\odot}$ neutron star to 
be $11.82\leqslant R_{1.4}\leqslant13.72$ km.}

\end{abstract}
\maketitle
\newpage

\section{Introduction}  
\label{intro}
The physics of dense matter relevant to neutron stars (NSs) is poorly
understood till date \cite{book_Haensel2007}. Neutron stars are made
of incredibly dense matter reaching densities up to few times the
nuclear saturation density ($\rho_0\sim0.16$ fm$^{-3}$) in the core
region. The NS structure depends predominantly on the nuclear equation
of state (EoS). Due to the lack of detailed knowledge of the nuclear
interactions at densities typical of the NS interior, many theoretical
models of nuclear EoS have been proposed.  Matter at supra nuclear
densities, as encountered in the NS interior, is difficult to access
in terrestrial experiments.  Inputs from astrophysical observations
are, therefore,  crucial in constraining the dense matter EoS.  Currently, the most
stringent constraint comes from the observation of NS with $\sim2~{\rm
M}_\odot$ \cite{Antoniadis2013,Arzoumanian} which sets a lower limit
for the maximum mass to be predicted by an EoS.

As NSs are massive and compact astrophysical objects, the coalescence
of binary NS systems is one of the most promising sources of
gravitational waves (GWs) observable by ground-based detectors
\cite{Taylor82,Drever83,Hawking87,Brillet89, Hough89,Cutler1993}. The GW
signals emitted during a NS merger depends on the behavior of {neutron}
star matter at high densities \cite{Faber2009,Duez2010}. Therefore,
its detection opens the possibility to constrain the nuclear matter
parameters characterizing the EoS.  A significant signature carried by GWs is
the tidal deformability (polarizability) of the NS and it is well
explored analytically \cite{Flanagan2008,Hinderer2008,Damour2009,
Binnington2009,Hinderer2010}. In a coalescing binary NS system, during
the last stage of inspiral,  {each NS} develops a mass {quadrupole} due
to the extremely strong tidal gravitational field induced by the other
{NS forming} the binary. The dimensionless tidal deformability $\Lambda$
describes the degree of deformation of a {NS} due to the tidal field of
the companion {NS} and is sensitive to the nature of the EoS.

In August 2017 the Advanced LIGO and Advanced Virgo gravitational-wave observatories 
detected GWs emitted from a binary NS inspiral for the first time
\cite{GW170817}. Remarkably, this discovery opened a new window
in the field of multi-messenger astronomy and nuclear physics, which
revealed {the} potential to directly probe the physics of {NSs} and {of}
the synthesis of heavy elements in the rapid neutron-capture process
({\it r}-process) \cite{GW170817_kilonova,Villar2017}. 
{The analysis of GW170817 data has allowed to put an
upper bound on {the} NSs combined dimensionless tidal deformability with 
90 \% confidence, using spin magnitudes consistent with the observed
neutron star population. In the analysis, results for both a high-spin and a low-spin
prior have been obtained to the same level of confidence. In our
study we will consider the constraints set by the low-spin prior
because they
are  consistent with the masses of all known binary neutron star
systems.
{This prior predicts that the combined dimensionless tidal deformability of
  the NS merger  is $\tilde\Lambda\le 800$. {In \cite{SDe2018} a
  reanalysis of the gravitational-wave observations of the binary
  neutron star merger GW170817 has been done assuming the same EoS for
both stars and  supplementing the gravitation-wave observation with information on the  source
location and distance from electromagnetic observations.}
 For the low spin prior these authors have obtained the
constraint $\tilde\Lambda\le 1000$. On the other end, the
investigation of the  UV/optical/infrared counterpart of GW170817 with
kilonova models {and complemented with numerical relativity results} in \cite{Radice2018} has set a lower bound on
$\tilde\Lambda$, i.e. $\tilde\Lambda>400$. {It should, however, be
  mentioned that this lower bound was obtained from 29 merger simulations
  covering several masses such that $q\gtrsim 0.85$ \cite{Radice2018a} and restricted to three models of nuclear
  matter, one
including also the $\Lambda$-hyperon.}}
{We show that these bounds on  the $\tilde \Lambda$ can be employed to deduce the 
respective bounds on the tidal deformability of a NS with mass $1.4~ {\rm M}_\odot$.}

Studies of the correlations between nuclear
matter parameters and the tidal deformability, based on a few selected
relativistic mean field models, have shown that measurements of the
{latter} can constrain the high density behavior of the nuclear symmetry
energy \cite{Fattoyev2013} as well as put bounds on the value of neutron
skin thickness \cite{Fattoyev2017}.  
These preliminary {studies need}
to be {validated} further using a more diverse set of 
models {for the nuclear EoS.} 
In {earlier studies} it was found that correlations between
the various properties of NS and nuclear matter EoS parameters {are}
significantly affected when a more diverse set of models are employed
\cite{Alam2015,Fortin2016}.
{Recently, astrophysical observations of NS,
in particular, the maximum mass, the radius of a canonical 1.4$~ {\rm M}_\odot$
NS, and the tidal deformability, have been used to constrain various
parameters of the EoS \cite{Zhang2018}. However, within their assumptions,
they found that the tidal deformability obtained from GW170817
is not very restrictive.}

{The present communication is an attempt, in view of the recent observation GW170817, 
to further explore the dependence
of the tidal deformability on the various nuclear matter parameters
describing the EoS.}  We study
the correlations of {the} tidal deformability parameter with the different
several nuclear matter parameters associated with a EoS by employing a representative set of relativistic
mean field (RMF) models and of Skyrme Hartree-Fock (SHF) models. The {considered}
EoS parameters are the nuclear matter incompressibility
coefficient, the symmetry energy coefficient and their derivatives at
the saturation density. We also study the model dependence of the Love
number $k_2$ which plays a crucial role in determining the value of
tidal deformability.

The paper is organized as follows. In Sec. II, we briefly outline the 
procedure for computing the tidal deformability and also define the
various nuclear matter parameters which can be calculated for a given 
EoS. In Sec. III we present the EoSs for our 
representative set of RMF and SHF models and use them
to calculate the tidal deformability and the Love number over 
a wide range of NS masses. The main results for the correlations of 
{the} tidal deformability, Love number and NS radius with different nuclear matter
parameters are discussed in Sec. IV. Finally the conclusions are drawn in Sec. V.

{\it Conventions:} We have taken the value of $\it{G = c = 1}$ throughout the
manuscript.

\section{Framework}
       In this section, we outline the expressions required to compute the 
       tidal deformability for a given EoS. We also define the various nuclear 
       matter parameters that characterizes the EoS. 
       
\subsection{Tidal deformability}
The tidal deformability parameter $\lambda$ is defined as \cite{Flanagan2008,Hinderer2008,
Hinderer2010,Damour2012},
\begin{equation}
Q_{ij}=-\lambda {\cal E}_{ij},
\end{equation}
where $Q_{ij}$ is the induced {quadrupole} moment of a star in a binary due to the static external tidal 
field ${\cal E}_{ij}$ of the companion star. The parameter $\lambda$ can be expressed in terms of the 
dimensionless {quadrupole} tidal Love  number $k_2$ as
\begin{equation}
\label{eq2}
\lambda = \frac{2}{3}k_2R^5,
\end{equation}
where $R$ is the radius of the NS. The value of $k_2$ is typically in 
the range $\simeq 0.05-0.15$ \cite{Hinderer2008,Hinderer2010,Postnikov2010} for NSs and depends on the stellar 
structure. This quantity can be calculated using the following expression \cite{Hinderer2008}
\bea
&& k_2 = \frac{8C^5}{5}\left(1-2C\right)^2
\left[2+2C\left(y_R-1\right)-y_R\right]\times \\
&&\bigg\{2C\left(6-3 y_R+3 C(5y_R-8)\right)\nn \\
&&+4C^3\left[13-11y_R+C(3 y_R-2)+2
C^2(1+y_R)\right] \nn \\
&& ~ ~
+3(1-2C)^2\left[2-y_R+2C(y_R-1)\right]\log\left(1-2C\right)\bigg\}^{-1},\nn
\eea
where $C$ $(\equiv m/R)$ is the compactness parameter of the star of
mass $m$.  The quantity $y_R$ $(\equiv y(R))$ can be obtained by solving
the following differential equation
\bea
r \frac{d y(r)}{dr} + {y(r)}^2 + y(r) F(r) + r^2 Q(r) = 0
\label{TidalLove2} ,
\eea
with
\bea
F(r) = \frac{r-4 \pi r^3 \left( \epsilon(r) - p(r)\right) }{r-2
m(r)},
\eea
\bea
\nonumber Q(r) &=& \frac{4 \pi r \left(5 \epsilon(r) +9 p(r) +
\frac{\epsilon(r) + p(r)}{\partial p(r)/\partial
\epsilon(r)} - \frac{6}{4 \pi r^2}\right)}{r-2m(r)} \\
&-&  4\left[\frac{m(r) + 4 \pi r^3
p(r)}{r^2\left(1-2m(r)/r\right)}\right]^2 \ .
\eea
{In the previous equations,} $m(r)$ is mass enclosed within the radius
$r$, and  $\epsilon(r)$ and $p(r)$ are, respectively,  the energy density and pressure in terms of radial coordinate $r$
of a star. These quantities are calculated within the
nuclear matter 
model chosen to describe the stellar EoS.
For a given EoS, Eq.(\ref{TidalLove2}) can be integrated together with the Tolman-Oppenheimer-Volkoff 
equations \cite{Weinberg} with the boundary conditions 
$y(0) = 2$, $p(0)\!=\!p_{c}$ and $m(0)\!=\!0$,{ where} $y(0)$, $p_c$ and $m(0)$ are the dimensionless quantity, pressure and mass 
at the center of the NS, respectively. One can then define the dimensionless tidal deformability: 
$\Lambda = \frac{2}{3}k_2 C^{-5}$. 
The tidal deformabilities of the NSs present in the binary neutron star system can be combined to yield 
the weighted average as,  
{\bea
\label{lambr}
\tilde \Lambda = \frac{16}{13} \frac{(12 q + 1) \Lambda_1 
+ (12 + q  ) q^4 \Lambda_2}{( 1 + q)^5},
\eea
where $\Lambda_1$ and $\Lambda_2$ are the individual tidal deformabilities corresponding 
to the two components in the NS binary with masses $m_1$ and $m_2$, respectively \cite{Flanagan2008,Favata2014}
with $q=m_2/m_1 < 1$.}       
       
\subsection{The nuclear matter parameters}
The energy per nucleon at a given density  $\rho=\rho_n+\rho_p$ with $\rho_n$ and $\rho_p$ the neutron and proton
densities, respectively, and asymmetry $\delta=(\rho_n-\rho_p)/\rho$, can be decomposed, to a good approximation, 
into the EoS for symmetric nuclear matter $e(\rho,0)$, and the density dependent symmetry energy coefficient $S(\rho)$:
\bea
 e(\rho,\delta)\simeq e(\rho,0)+S(\rho)\delta^2.
 \label{eq:eden}
\eea 
Expanding the isoscalar contribution until {third} order and the isovector 
until {second} order we obtain for the 
isoscalar part $e(\rho,0)$:
\begin{small}
\bea
e(\rho,0)= e(\rho_0)+\frac{K_0}{2} x^2+\frac{Q_0}{6} x^3 + \mathcal{O}(x^4)
\label{eq:E}\hspace{+13pt}
\eea
\end{small}
and for the isovector part $S(\rho)$:
\begin{small}
\begin{eqnarray}
S(\rho)=J_0+L_0 x +\frac{K_{\rm sym,0}}{2}  x^2 + \mathcal{O}(x^3).
\label{eq:S}
\end{eqnarray}
\end{small}
where $x=\frac{\rho-\rho_0}{3\rho_0}$ and $J_0=S(\rho_0)$ is the symmetry energy at the saturation density. The incompressibility $K_0$, the 
skewness coefficient $Q_0$, the symmetry energy slope $L_0$, and its curvature $K_{\rm sym,0}$ evaluated at saturation density
are defined in, e.g.,  Ref. \cite{Vidana2009}.
The slope of the incompressibility, $M_0$, at saturation density is defined as \cite{Alam2015},
\bea
M_0=12K_0 + Q_0 .
\eea
In the section \ref{results} we shall consider the correlations of the tidal deformability of NS
with the various nuclear matter parameters of the EoS: $K_0$, $Q_0$, $M_0$, $J_0$, $L_0$, $K_{\rm sym,0}$.

\section{Equation of state and tidal deformability}
In the present section we introduce a set of relativistic and non-relativistic
nuclear models that are constrained by the bulk properties of finite nuclei 
and the observed lower bound on the NS maximum mass. For these models we show how the 
tidal deformability and Love number behave over a wide range of NS masses.

\subsection{Nuclear matter equation of state}
\label{eos}
The correlations of the properties of neutron stars with  
the various nuclear matter parameters of the EoS are studied using a set of  
eighteen relativistic and twenty-four non-relativistic
nuclear models. These models have been
employed for the study of finite nuclei and NS properties. Our set of
models are based on RMF and SHF frameworks. The {employed} RMF models
are BSR2, BSR3, BSR6 \cite{Dhiman2007,Agrawal10},FSU2
\cite{Chen14}, GM1 \cite{Glendenning91}, NL3 \cite{Lalazissis97},
NL$3{\sigma\rho}4$, NL$3{\sigma\rho}6$ \cite{Pais16}, NL$3{\omega
\rho}02$ \cite{Horowitz01}, NL$3{\omega \rho}03$ \cite{Carriere03}, TM1
\cite{Sugahara94}, and TM1-2 \cite{Providencia13}and DD2 \cite{Typel10},
DDH$\delta$ \cite{Gaitanos04}, DDH$\delta$Mod \cite{Ducoin11}, DDME1
\cite{Niksic02}, DDME2 \cite{Lalazissis05}, and TW \cite{Typel99}. The {considered}
SHF models are the SKa, SKb \cite{Kohler76},
SkI2, SkI3, SkI4, SkI5 \cite{Reinhard95}, SkI6 \cite{Nazarewicz96},
Sly2, Sly9 \cite{ChabanatPhd}, Sly230a \cite{Chabanat97}, Sly4
\cite{Chabanat98}, SkMP \cite{Bennour89}, SKOp \cite{Reinhard99}, KDE0V1
\cite{Agrawal05}, SK255, SK272 \cite{Agrawal03}, Rs \cite{Friedrich86},
BSk20, BSk21 \cite{Goriely10}, BSk22, BSk23, BSk24, BSk25, and BSk26
\cite{Goriely13}. 
{The values of the nuclear matter properties, such as, $K_0$, $Q_0$, $M_0$, $J_0$, $L_0$ 
and $K_{\rm sym,0}$ vary over a wide range for our representative set of EoSs 
as can be seen from the supplementary material of
Ref. \cite{Alam2016}.
As the mass of the stars in the GW170817 binary is
{1.6}$~ {\rm M}_\odot$ or smaller, we only consider nucleonic degrees of
freedom. However, a NS with a mass of 1.6 $~ {\rm M}_\odot$ could have
non-nucleonic degrees of freedom \cite{Dhiman2007,Fortin2017}.}

          The EoSs {considered} for all the models are consistent with the
observational constraint provided by the existence of $2~ {\rm M}_\odot$
NS \cite{Fortin2016,Alam2016}. Moreover, the {considered} SHF models do not
become acausal for masses below $2~ {\rm M}_\odot$. We have taken unified
inner-crust core EoS for all the models \cite{Fortin2016} and the EoS of
Baym-Pethick-Sutherland \cite{Baym1971} is used for the outer
crust.

In Fig. \ref{fig1}, we plot for NS matter the
variation of pressure ($p$) with {the} energy density ($\varepsilon$)
in the left panel and the variation of $dp/d\varepsilon$ with the
baryon number density in the right panel for our representative set of
models. The black circles denote the central density corresponding to the
NS maximum mass for each EoS. The dashed line indicates the causality
limit (i.e. $dp/d\varepsilon=1$). The values of $dp/d\varepsilon$ for
SHF models are larger at higher densities ($\rho \gg \rho_0$) than
those for the RMF models.
{The  maximum mass NS configurations of all models studied 
are within the causality limit except for BSk20 and
BSk26  EoSs, which are marginally acausal.}
\begin{figure}[htb]
 \centering
\begin{tabular}{c}
    \includegraphics[width=.45\textwidth,angle=0]{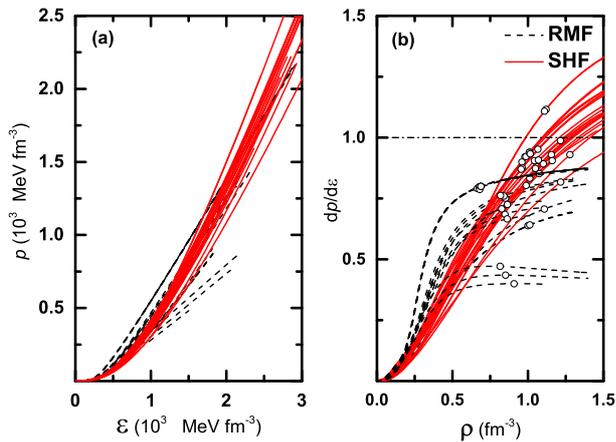}
    \end{tabular}
    \caption{(Color online) Plots for the  (a) pressure $p$ as a function of {the} energy density, 
                            and (b) $dp/d\varepsilon$ as a function of {the} baryonic number density 
                            for beta equilibrated NS matter obtained using a representative 
                            set of RMF (black dashed lines) and SHF models (red lines). 
                            The circles in right panel correspond to the central
                            densities and the slopes $dp/d\varepsilon$ at the maximum NS mass for 
                            each of the EoS. The BSk20 and BSk26 EoSs are marginally acausal 
                            at the NS maximum masses $\sim 2.2~{\rm M}_\odot$ 
                            \cite{Fortin2016,Alam2016}.        
            }
    \label{fig1}
\end{figure}

\subsection{Dependence of the tidal deformability on the equation of state}
      One of the main focus of the present work is to study the sensitivity of the tidal deformability
to the properties of nuclear matter at saturation density. To facilitate our discussions in the next section,
in Fig. \ref{fig2} the dimensionless tidal deformability $\Lambda$ (left) and 
tidal Love number $k_2$ (right) obtained for our set of EoSs are plotted as a function of the NS mass. 
The values of $k_2$ show a noticeable spread across the various models. For instance, 
at  $1.4~ {\rm M}_{\odot}$, the values of $k_2$ are in the range of {$0.07$ to $0.11$}. 
For the smaller masses the spread in $k_2$ is larger for the SHF models, but for the larger 
masses RMF models give on average larger values of  $k_2$. {One can also see from Fig. 1 
of reference \cite{Alam2016} that the RMF models predict larger radii, in particular, for 
{large} NS masses. Consequently, the parameter $\Lambda$ tends to be larger for the RMF 
models than for the SHF models. In the following, we will examine the dependence of $\Lambda$ 
on both $k_2$ and $R$ in detail.}
\begin{figure}[htb]
\centering
\begin{tabular}{c}
     \includegraphics[width=.45\textwidth,angle=0]{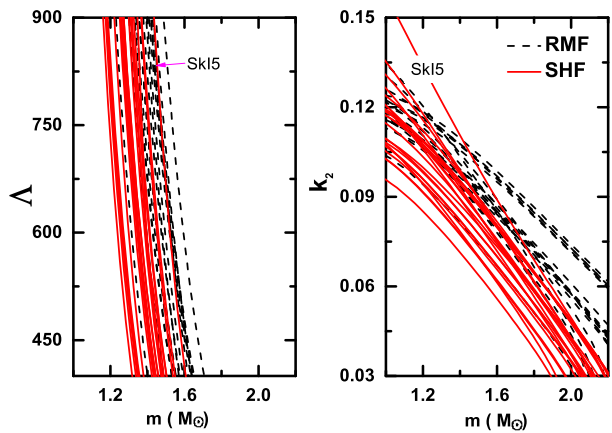}
     \end{tabular}
     \caption{(Color online) (a) Tidal deformability $\Lambda$ and (b) the Love
     number $k_2$ as a function of the NS mass (m) for a representative
     set of relativistic and non-relativistic models. The SHF model,
     SkI5, displays markedly different behavior for $\Lambda$ as well
     as for $k_2$.}
     \label{fig2}
\end{figure}

   In Fig. \ref{fig3} we plot the tidal deformabilities
in the phase space of $\Lambda_1$ and $\Lambda_2$ associated,
respectively, with the
high-mass $m_1$ and the low-mass $m_2$ components of the binary, for all
the {considered} RMF and SHF models. The curves corresponding to every
EoS {are obtained} by varying the high mass ($m_1$) independently in the
range $1.365<m/{\rm M}_\odot < 1.60$ obtained for GW170817 whereas the low mass
($m_2$) is determined by keeping the chirp mass $\mathcal{M} =(m_1 m_2)^{3/5} 
(m_1+m_2)^{-1/5}$ fixed
at the observed value $1.188~{\rm M}_\odot$ \cite{GW170817}. The
dot-dot-dashed and the dot lines represent, respectively,  the 90\% and 50\% confidence limits obtained 
from the GW170817 for the low spin priors.
One can
note that the $90\%$ confidence limit suggests
that SkI5 and the family of models NL3X and TM1X are ruled out except
for NL$3{\omega \rho}03$.  For the SkI5 the values of $M_0$ and $L_0$
are 2745 MeV and 129 MeV, respectively. 
{For NL3X family the value of $M_0$ is larger than 3400 MeV and $L_0$
is in the range of  55-70 MeV except for the base
model NL3. Whereas, for TM1X family the value of $M_0\sim3100$ MeV and $L_0\sim110$ MeV.}
{This indicates that very high value of $M_0$ and/or $L_0$
may not be favored by GW170817.}  
\begin{figure} [htb]
 \centering
\begin{tabular}{c}
    \includegraphics[width=.45\textwidth,angle=0]{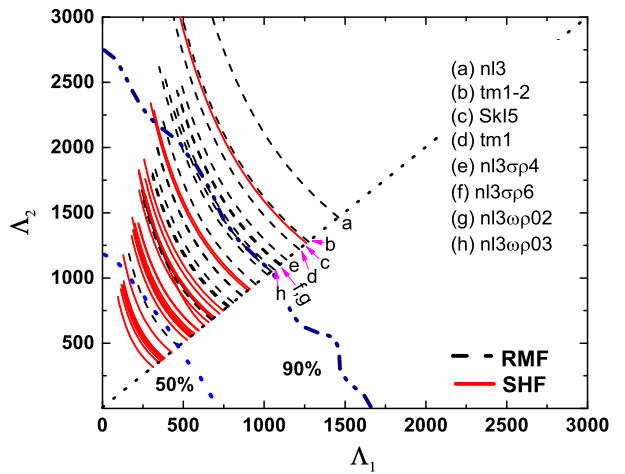}
\end{tabular}
  \caption{(Color online) Tidal deformability parameters for the case of high mass {($\Lambda_1$)}
           and low mass {($\Lambda_2$)} components of the observed GW170817. The 
           90\%(dot-dot-dashed) and 50\% (dot) confidence lines  
           are taken from Ref. \cite{GW170817} corresponding to the low spin priors.
           }
\label{fig3}
\end{figure}

\section{Results and Discussions}
\label{results}
  In the present section, we study the correlations of the tidal deformability
$\Lambda$, the Love number $k_2$ and the radius of NSs $R$ with  various nuclear matter 
parameters. As already mentioned in Sec. \ref{intro}, we consider 
the constraints from the properties of the binary neutron star that satisfy 
the low spin prior \cite{GW170817}. In our analysis, the correlation between a 
pair of quantities is quantified in terms of Pearson’s correlation coefficient, 
denoted as $\mathcal{R}$ \cite{Brandt97}. The magnitude of $\mathcal{R}$ is at 
most unity indicating that the pair of quantities is 
completely correlated to each other. For $|\mathcal{R}| < 0.5$, the correlations 
are usually said to be weak.

    We calculate the values of the coefficients for the correlation of
$\Lambda$, $k_2$ and $R$ with the nuclear matter saturation parameters 
$K_0$, $Q_0$, $M_0$, $J_0$, $L_0$, $K_{\rm sym,0}$ and {with 
several linear combinations of two parameters}, in particular with 
$K_0+ \alpha L_0$, $M_0+ \beta L_0$ and $M_0+ \eta K_{\rm sym,0}$. 
The values of $\alpha$, $\beta$ and $\eta$ are obtained {so} that, 
{for each NS mass}, they yield optimum correlations. 
Our correlation systematics is determined for NS masses in the 
range of $1.2-1.6~{\rm M}_\odot$, since, for the low spin prior analysis, 
these masses are close to the ones involved in the GW170817 event.
{The results for the values of the $\mathcal{R}$ obtained for the correlation 
of $\Lambda$, $k_2$ and $R$ with individual nuclear matter parameters are presented
in Table \ref{tab1}. The Table \ref{tab2} contains the results obtained 
using the linear combinations of the nuclear matter parameters.
The Fig. \ref{fig4} is the 
pictorial representation of the results presented in Tables \ref{tab1} and \ref{tab2}. 
\begin{table} [htb]
\caption{\label{tab1} {The Pearson correlation coefficients, $\mathcal{R}$ obtained
                 for the correlations between various
                 NS and nuclear matter properties. The values of tidal
                 deformability $\Lambda$, radius $R$ and the Love number
                 $k_2$ are evaluated for the NS masses $1.2-1.6$
                 $~{\rm M}_\odot$. The nuclear matter incompressibility
                 $K_0$, skewness $Q_0$, slope of incompressibility
                 $M_0$, symmetry energy $J_0$, slope of symmetry energy
                 $L_0$ and the curvature parameters $K_{\rm sym,0}$ at saturation density.}}
\begin{ruledtabular}
\begin{tabular}{ccccccc}
                &       $K_0$   &       $Q_0$   &       $M_0$   &       $J_0$   &       $L_0$   &      $K_{{\rm sym},0}$ \\
\hline                
$\Lambda_{1.2}$	&	0.68	&	0.46	&	0.68	&	0.58	&	0.81	&	0.76	\\
$\Lambda_{1.3}$	&	0.69	&	0.51	&	0.72	&	0.56	&	0.76	&	0.74	\\
$\Lambda_{1.4}$	&	0.70	&	0.57	&	0.76	&	0.53	&	0.71	&	0.71	\\
$\Lambda_{1.5}$	&	0.71	&	0.62	&	0.80	&	0.50	&	0.65	&	0.68	\\
$\Lambda_{1.6}$	&	0.71	&	0.66	&	0.82	&	0.46	&	0.59	&	0.64	\\
\hline													
$R_{1.2}$	&	0.65	&	0.48	&	0.67	&	0.65	&	0.82	&	0.70	\\
$R_{1.3}$	&	0.66	&	0.51	&	0.70	&	0.62	&	0.79	&	0.70	\\
$R_{1.4}$	&	0.67	&	0.54	&	0.72	&	0.59	&	0.75	&	0.69	\\
$R_{1.5}$	&	0.68	&	0.57	&	0.75	&	0.56	&	0.72	&	0.68	\\
$R_{1.6}$	&	0.68	&	0.60	&	0.77	&	0.53	&	0.68	&	0.66	\\
\hline											
$k_{2,1.2}$	&	0.57	&	0.34	&	0.54	&      -0.03	&	0.44	&	0.79	\\
$k_{2,1.3}$	&	0.62	&	0.47	&	0.65	&	0.02	&	0.43	&	0.76	\\
$k_{2,1.4}$	&	0.64	&	0.55	&	0.72	&	0.05	&	0.39	&	0.72	\\
$k_{2,1.5}$	&	0.65	&	0.63	&	0.77	&	0.08	&	0.36	&	0.66	\\
$k_{2,1.6}$	&	0.58	&	0.59	&	0.71	&	0.06	&	0.26	&	0.57	\\
\end{tabular}   
\end{ruledtabular}
\end{table}
\begin{figure*} 
\begin{tabular}{ccc}
    \includegraphics[width=0.30\textwidth,angle=0]{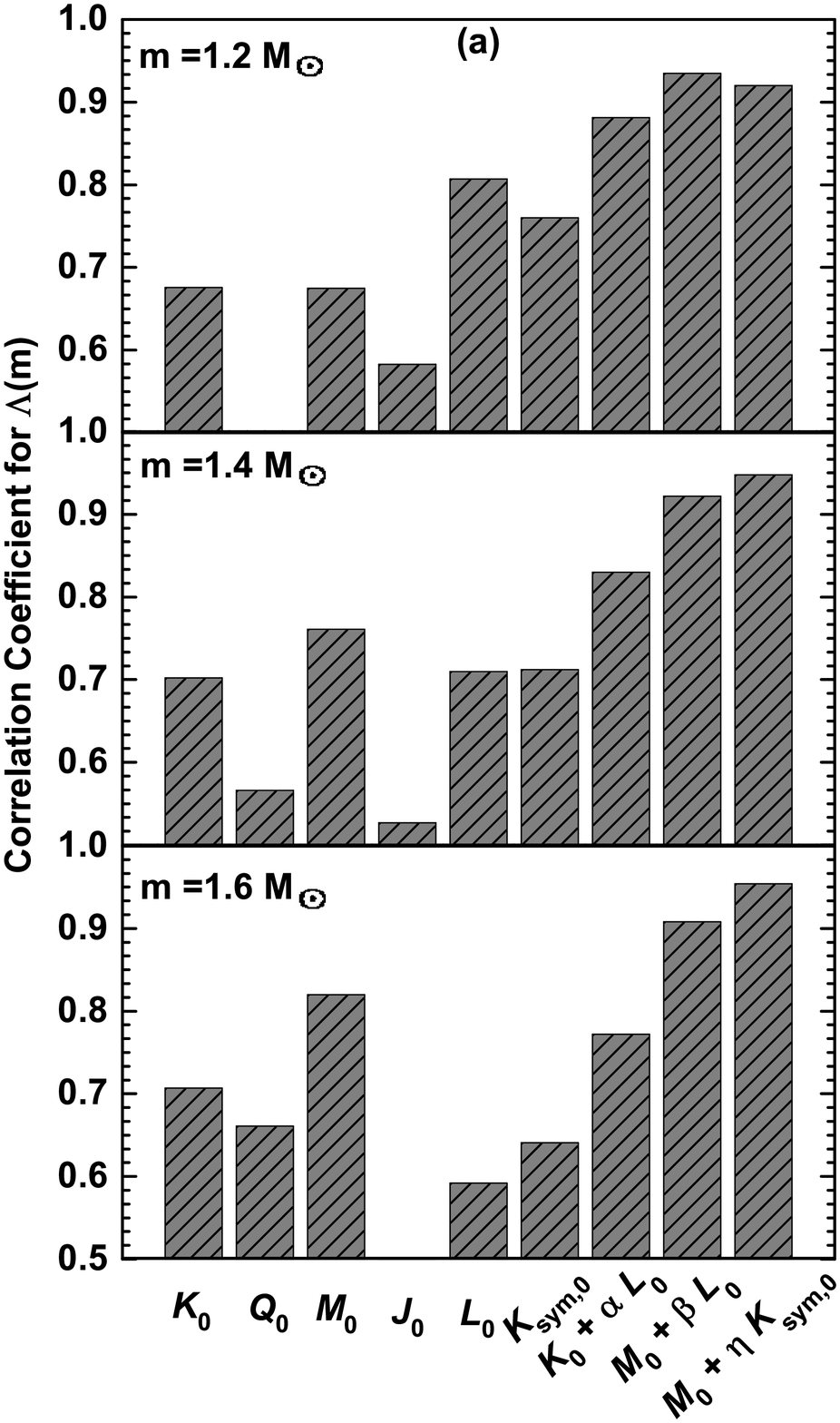} &
    \includegraphics[width=0.30\textwidth,angle=0]{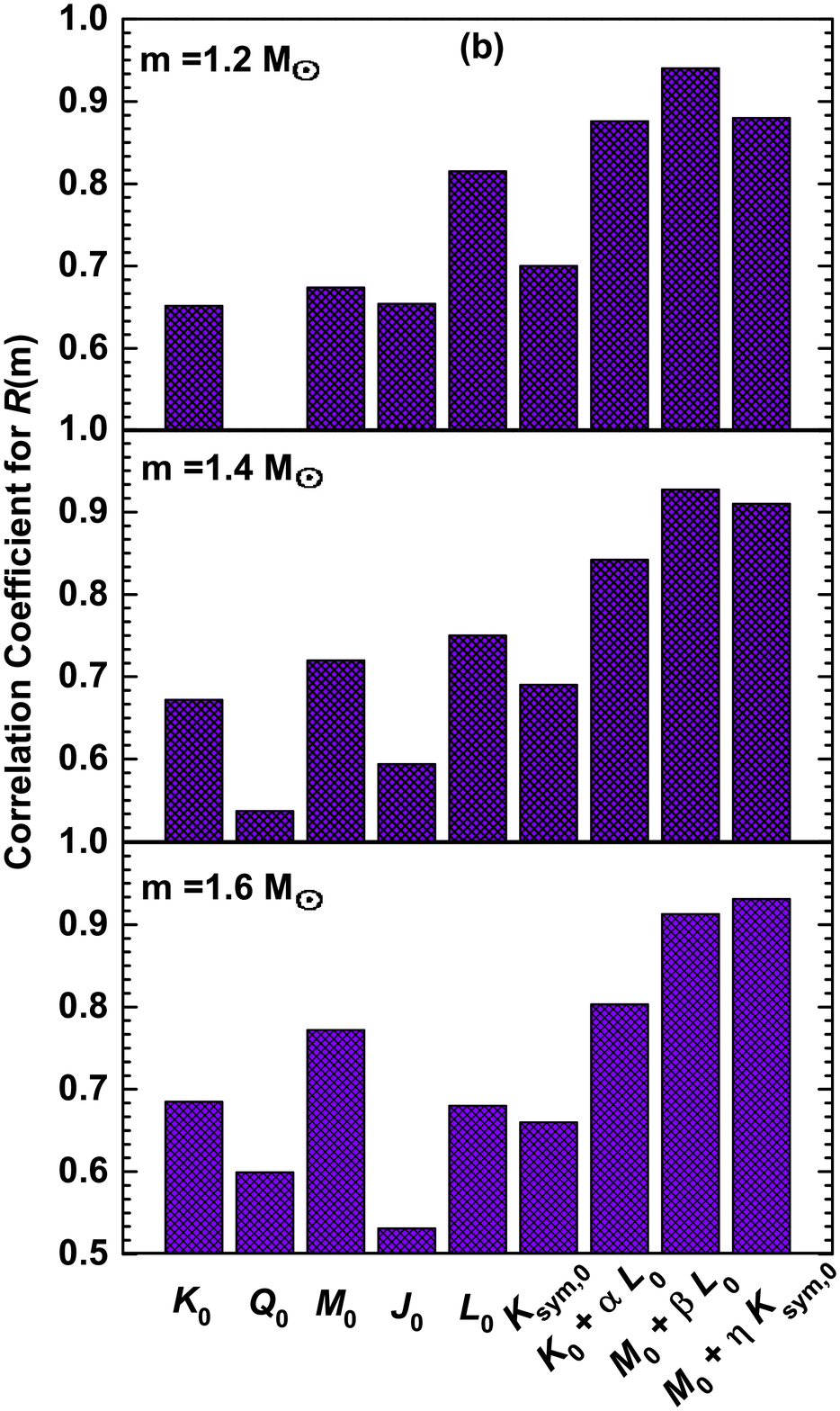} & 
     \includegraphics[width=0.30\textwidth,angle=0]{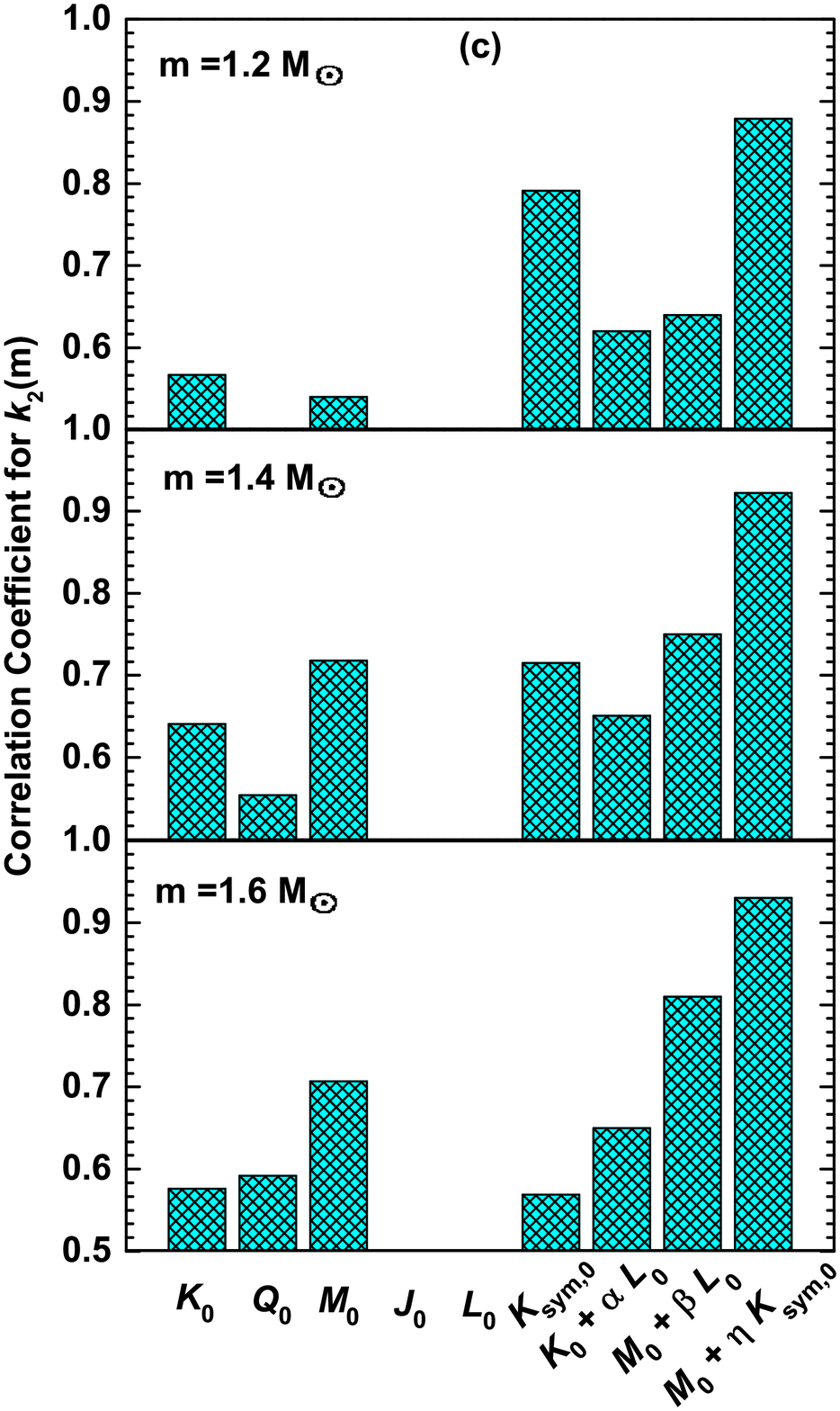}  
\end{tabular}
  \caption{(Color online) Correlation coefficients $\mathcal{R}$ for (a-c) the tidal deformability $\Lambda$, (d-f) the
                           radius $R$, and (g-i) the Love number $k_2$ with different individual nuclear matter parameters
                           as well as with some selected linear combinations of them
                           obtained for the {NS} masses $1.2~ {\rm M}_{\odot}$ (top), $1.4 ~ {\rm
                           M}_{\odot}$ (middle) and $1.6~ {\rm M}_{\odot}$ (bottom). Results are
                           plotted only for the cases with $\mathcal{R}> 0.5$ (see tables \ref{tab1} and
                           \ref{tab2} for details).}
 \label{fig4}
\end{figure*}
Only the 
cases with the correlation coefficients $\mathcal{R}>0.5$ are displayed.
We see from Table \ref{tab1} 
that for most of the cases, individual EoS parameters seem to be weakly or moderately correlated with 
the $\Lambda$, $k_2$ and $R$. Exceptionally, the $\Lambda$ and $R$ are strongly correlated
with the individual nuclear matter parameters $L_0$ and $M_0$ for the NS masses 1.2$~ {\rm M}_\odot$ and 1.6$~ {\rm M}_\odot$,
respectively.} Let us point out that the correlation between the {radius of low 
mass NSs} and the neutron skin of {$^{208}$Pb}, {which is itself correlated 
with $L_0$}, was first discussed in \cite{Horowitz01,horowitz2001}.
It is seen from Table \ref{tab2}, the $\Lambda$ and $R$ are strongly
correlated with $M_0+\beta L_0$ and $M_0+ \eta K_{\rm sym,0}$ over a 
wide range of NS masses considered: the values 
of $\mathcal{R}$ of the order of 0.9. The Love number $k_2$ is strongly 
correlated with $M_0+ \eta K_{\rm sym,0}$. 
The values of $\alpha$, $\beta$ and $\eta$ decrease monotonically with 
the NS mass. {This indicates that the density dependence of symmetry 
energy is less important in determining the values of $\Lambda$ and $R$ 
at higher NS masses.} The mass dependence 
of $\alpha$, $\beta$ and $\eta$ is discussed in some detail in the 
Appendix \ref{apd1}, where, in particular, an exponential dependence of these
parameters on the NS mass
is proposed.
\begin{table} 
\caption{\label{tab2} {The values of the coefficients $\mathcal{R}$ obtained 
                      for the correlations of $\Lambda$, $R$ and $k_2$ with various linear combinations of 
                      EoS parameters. The calculations are performed for the NS masses
                      $1.2-1.6~{\rm M}_\odot$.}}
\begin{ruledtabular}
\begin{tabular}{c|cc|cc|cc}
   &  \multicolumn{2}{c|}{$K_0+\alpha L_0$} &\multicolumn{2}{c|}{$M_0+\beta L_0$}&\multicolumn{2}{c}{$M_0+\eta K_{\rm sym,0}$}\\

                 &$\mathcal{R}$ & $\alpha$ & $\mathcal{R}$ & $\beta$ & $\mathcal{R}$ & $\eta$ \\ 
\hline                 
$\Lambda_{1.2}$  &   0.88 &  1.16 &    0.94   &  21.22 &     0.92  &   6.34   \\
$\Lambda_{1.3}$  &   0.86 &  0.93 &    0.93   &  17.05 &     0.94  &   5.55   \\
$\Lambda_{1.4}$  &   0.83 &  0.74 &    0.92   &  13.68 &     0.95  &   4.83   \\
$\Lambda_{1.5}$  &   0.80 &  0.59 &    0.92   &  10.91 &     0.95  &   4.18   \\
$\Lambda_{1.6}$  &   0.77 &  0.45 &    0.91   &  8.54  &     0.95  &   3.62   \\
\hline
$R_{1.2}$        &  0.88    & 1.33	&    0.94   &   21.75 &    0.88 & 5.64      \\
$R_{1.3}$ 	 &  0.86    & 1.14	&    0.93   &   19.07 &    0.90 & 5.33      \\
$R_{1.4}$ 	 &  0.84    & 0.98	&    0.93   &   16.62 &    0.91 & 5.00      \\
$R_{1.5}$ 	 &  0.82    & 0.84	&    0.92   &   14.38 &    0.92 & 4.65      \\
$R_{1.6}$ 	 &  0.80    & 0.71	&    0.91   &   12.32 &    0.93 & 4.31       \\
\hline
$k_{2,1.2}$	&  0.62    &  0.40  &  0.64  & 11.18  &  0.88  &  9.15      \\
$k_{2,1.3}$	&  0.64    &  0.25  &  0.70  & 7.22   &  0.91  &  6.83      \\
$k_{2,1.4}$	&  0.65    &  0.16  &  0.75  & 4.81   &  0.92  &  5.31      \\
$k_{2,1.5}$	&  0.66    &  0.10  &  0.79  & 3.34   &  0.93  &  4.20      \\
$k_{2,1.6}$	&  0.65    &  0.04  &  0.81  & 2.14   &  0.93  &  3.52      \\
\end{tabular}
\end{ruledtabular}
\end{table}
As an example, in Fig.\ref{fig5} we plot $M_0+\beta L_0$ and 
$M_0+\eta K_{\rm sym,0}$ as a function of $k_2$ and $\Lambda$ for 
$1.4\,~ {\rm M}_{\odot}$ NS.
{Since, $\Lambda_{1.4}$ is not very well correlated individually
with $M_0$, $L_0$ and $K_{\rm sym,0}$, its strong correlation
with $M_0+\beta L_0$ and $M_0+\eta K_{\rm sym,0}$ is of 
particular importance.} 
 The values of the correlation coefficients given in the figure 
are obtained with the entire set of RMF and SHF models as presented in 
section \ref{eos}. In order to check the model dependence of the correlations, 
we have determined the correlation coefficients for the sets of RMF 
and SHF models separately. The results are given in Table \ref{tab3} 
which indicate that the model dependence is only marginal.

{The result for the correlations among $k_2$, $\Lambda$ and various nuclear matter 
properties as depicted in Fig. \ref{fig5} 
may be understood as follows.} In Ref. \cite{Alam2016}, 
it was shown that the NS radius $R$ is strongly correlated with a linear 
combination of $M_0$ and $L_0$ over a wide range of NS masses.
{This was attributed to the dependence of the pressure on $M_0$ and $L_0$ 
and to the empirical relation of the star radius with the pressure at 
several reference densities, e.g. $R \times p(\rho)^{-1/4}={\rm constant}$ for 
$\rho\sim 1.5 ~\rho_0$ and NS masses, $1-1.4~{\rm M}_\odot$, irrespective 
of the model \cite{Lattimer2000}.} 
\begin{figure} 
 \centering
\begin{tabular}{c}
    \includegraphics[width=.45\textwidth,angle=0]{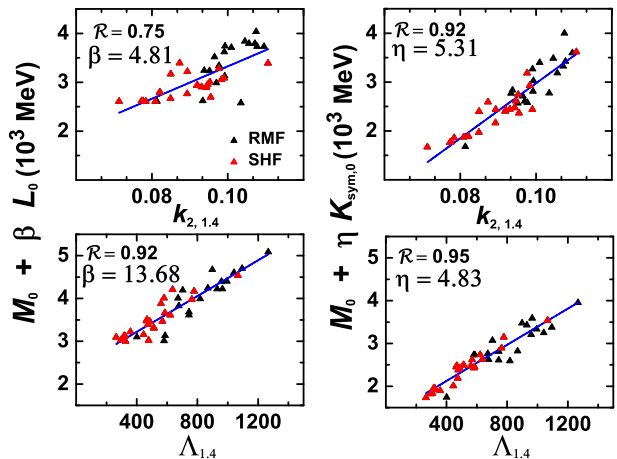}
    \end{tabular}
    \caption{(Color online) (a-b) The $M_0+\beta L_0$ and (c-d) $M_0+\eta K_{\rm sym,0}$ 
                             versus the tidal Love number $k_{2,1.4}$ 
                             (top panels) and dimensionless tidal deformability 
                             $\Lambda_{1.4}$ (bottom panels) for a  
                             $1.4 ~ {\rm M}_{\odot}$ NS, using a set of RMF and SHF models.}
    \label{fig5}
\end{figure}  

\begin{table}
\label{tab3}
\caption{\label{tab3} Values for the correlations coefficients for $\Lambda_{1.4}$ and $k_{2,1.4}$ with 
$M_0+ \beta L_0$ and $M_0 + \eta K_{\rm sym,0}$ obtained separately for the RMF and SHF models. The 
values of the correlation coefficients corresponding to all the  models (ALL) are also listed.}
\newcommand{\mc}[3]{\multicolumn{#1}{#2}{#3}}
\begin{ruledtabular}
\begin{tabular}[t]{l|c|c|c|c|c|c}
    & \mc{3}{c|}{$M_0+ \beta L_0$} & \mc{3}{c}{$M_0+ \eta K_{\rm sym,0}$}\\
    & \mc{1}{c}{RMF} & \mc{1}{c}{SHF} & \mc{1}{c|}{ALL} & \mc{1}{c}{RMF} & \mc{1}{c}{SHF} & \mc{1}{c}{ALL}\\
    \hline
$\Lambda_{1.4}$	&	0.92	&	0.90	&	0.92		&	0.88		&	0.97		&	0.95	\\
$k_{2,1.4}$	&	0.72	&	0.68	&	0.75		&	0.89		&	0.91		&	0.92	\\
\end{tabular}
\end{ruledtabular}
\end{table}
The solid lines in Fig. \ref{fig5} are obtained using  linear regression. These linear regressions yield,   
{\bea
\label{eq_ml}
\frac{M_0}{{\rm MeV}} + 13.68 \frac{L_0}{{\rm MeV}} = (2.09 \pm 0.14) ~ \Lambda_{1.4}  \nonumber \\
                + (2383.12 \pm 96.42), \\
\label{eq_mksym}
\frac{M_0}{{\rm MeV}} + 4.83 \frac{K_{\rm sym,0}}{{\rm MeV}} = (2.11 \pm 0.11) ~ \Lambda_{1.4}  \nonumber \\
                   + (1278.13 \pm 77.76).
\eea}
We need to know the value of $\Lambda_{1.4}$ in order to exploit 
the correlations, as presented in Fig. \ref{fig5}, to 
estimate the values of nuclear matter properties at the saturation 
density.  

The GW170817 event provides the upper bound on $\tilde \Lambda$ as defined by
Eq.(\ref{lambr}). {For the low spin prior we have to consider
masses such that
$q=m_2/m_1 > 0.7$. We have calculated the $\tilde \Lambda$ using  
neutron star masses $m= 1.4,\, 1.17,\, 1.6~ {\rm M}_\odot$, which
correspond to the canonical mass and the lower and upper mass limits
covered by the low spin prior analysis. The neutron star binary companion mass is 
determined from the chirp mass  
$\mathcal{M}=1.188 ~ {\rm M}_\odot$: $m= 1.17,\, 1.6~ {\rm M}_\odot$
are, respectively, $m_2$ and $m_1$ corresponding to $q=0.7$; for the
canonical mass we get $q=0.95$ with
$m_1=1.40~ {\rm M}_\odot$ and $m_2=1.33~ {\rm M}_\odot$.}
\begin{figure} 
 \centering
\begin{tabular}{c}
    \includegraphics[width=.45\textwidth,angle=0]{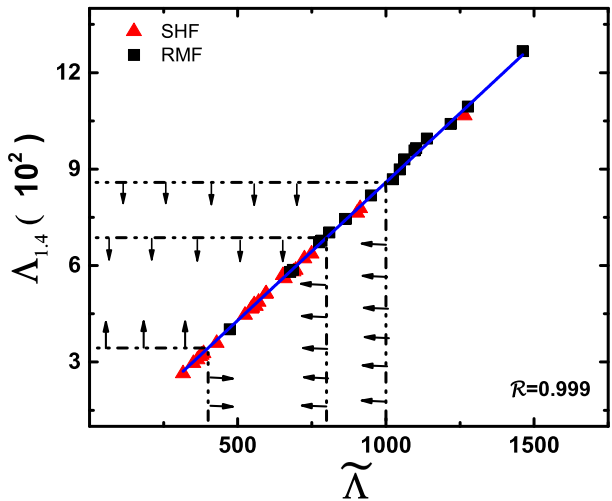}
    \end{tabular}
  \caption{(Color online) The tidal deformability $\Lambda_{1.4}$ verses the weighted average 
                          $\tilde \Lambda$ as defined in Eq.(\ref{lambr}) for all the RMF and 
                          SHF models. The solid line represents the best fit. The arrows pointing 
                          right and up indicate the lower bounds on $\tilde \Lambda$ and $\Lambda_{1.4}$,
                          respectively. The upper bounds on $\tilde \Lambda$ and $\Lambda_{1.4}$
                          are denoted by left and down arrows, respectively.}
\label{fig6_n}
\end{figure}
Fig. \ref{fig6_n} shows the variation of $\Lambda_{1.4}$ as a function of 
$\tilde \Lambda$ for all the RMF and SHF models. The correlation between this two quantities is very strong which enables 
us to express $\Lambda_{1.4}$ in terms of $\tilde \Lambda$ as 
$\Lambda_{1.4}=0.859 \times \tilde \Lambda$. {Similar studies were
performed for {the NS with mass} $m= 1.17$ and $\, 1.6~ {\rm M}_\odot$ and we have obtained
 $\Lambda_{1.17}=2.452 \times \tilde \Lambda$ and
$\Lambda_{1.6}=0.379 \times \tilde \Lambda$  with an equally strong
correlation. These relations should be compared with the prediction
from the expression proposed in \cite{SDe2018} 
\begin{equation}	
\Lambda_1=\frac{13}{16}\tilde \Lambda \frac{q^2
  (1+q)^4}{12q^2-11q+12}, \label{lambda_de}
\end{equation}
obtained by replacing 
\begin{equation}\Lambda_2= q^{-6}\Lambda_1
\label{tlam}
\end{equation}
 in the Eq.(\ref{lambr})
for $\tilde \Lambda$.  Eq.  (\ref{tlam}) was obtained assuming that the radii of the
stars with masses $1.17<m<1.6\, M_\odot$  are the same. 
Using  expression (\ref{lambda_de}),  we get
relations between $\Lambda_i$ and $\tilde\Lambda$ for $m_i=1.17, \,
1.4, \, 1.6\, M_\odot$ that coincide with ours within the first two digits. We have checked
that, {in most of the cases,} for our set of models the difference between the radii of stars with a
mass in that interval is not larger than $\sim 0.2$~km.

In the following, we want to constraint $M_0$ and $K_{\rm sym,0}$.
We will consider the limits imposed on $\Lambda_{1.4}$. This choice is
justified because according to the analysis done in \cite{SDe2018,LIGO2} the
limits obtained for $\tilde \Lambda$ are $q$ dependent, and,
in particular, in  \cite{SDe2018}  if the double neutron star or galactic neutron star distributions are considered the
maximum $\tilde \Lambda$ value is obtained, respectively,
for $q>0.9$ ($q>0.8$).  For the lower limit the results of
\cite{Radice2018,Radice2018a} were determined for $q>0.85$.}
{A lower bound {of $\Lambda_{1.4}>344$ is set} by 
the UV/optical/infrared counterpart of GW170817 that imposes
$\tilde \Lambda>400$ \cite{Radice2018,Radice2018a}. 
Similarly, the gravitational-wave observations set an upper bound 
$\Lambda_{1.4} <687$ or 
$\Lambda_{1.4} <859$, respectively from the bounds $\tilde \Lambda<800$  \cite{GW170817} and $\tilde \Lambda <1000$ \cite{SDe2018}.}
In what follows, we will use these bounds on $\Lambda_{1.4}$ together with Eqs. (\ref{eq_ml} 
and \ref{eq_mksym}) to constrain the nuclear matter properties. 

\begin{figure} 
\centering
\begin{tabular}{c}
    \includegraphics[width=.45\textwidth,angle=0]{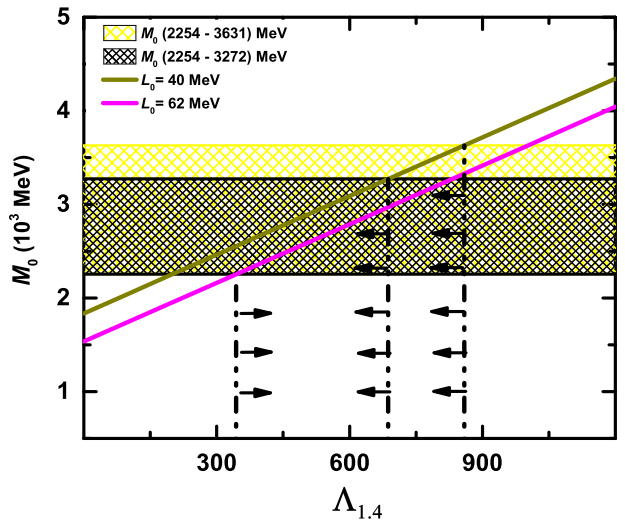}
    \end{tabular}
    \caption{(Color online) Plots for the  incompressibility slope parameter $M_0$ versus 
                          tidal deformability $\Lambda_{1.4}$  at fixed values of symmetry energy slope parameter 
                          $L_0$ 
                          (solid lines) obtained using Eq.(\ref{eq_ml}). The choices for 
                          the values of $L_0$ are
                          discussed in the text.
                          {The dot-dot-dashed lines represent the
                          bounds obtained in Fig. \ref{fig6_n}.}}
\label{fig6}
\end{figure}
    In Fig. \ref{fig6}, the slope of the incompressibility coefficient at the saturation density 
{$M_0$} is plotted as a function of $\Lambda_{1.4}$  for fixed values of $L_0$  using Eq. 
(\ref{eq_ml}). The limiting {values of} $L_0$ employed in the plot
correspond to 
$L_0=51\pm11$ MeV \cite{Lattimer2013}. This limit on $L_0$ in conjunction with the bounds on 
$\Lambda_{1.4}$, as discussed above, constrain the $M_0$ as listed in
Table \ref{tab4}. {As referred  before the
lower bound on $\tilde\Lambda$ set by \cite{Radice2018}  has several associated
uncertainties, and, therefore the lower bounds obtained for $M_0$
and $K_{\rm sym,0}$ suffer from these uncertainties. Notice, however, that
independently of the lower value of $\tilde\Lambda$ we always have $M_0> 1500$ (1800) MeV according to the constraints
imposed in $L_0$ in \cite{Lattimer2013} (\cite{Oertel2017}).}
In the same table {we also present} the values of $M_0$ obtained for $L_0=58.7\pm28.1$
MeV \cite{Oertel2017}. {These values of $L_0$} take into account 
terrestrial, theoretical and observational constraints. Our values of $M_0$ have a reasonable 
overlap with the values $M_0=(1800-2400)$ MeV obtained empirically in Ref. \cite{De15}. The value of $M_0$ 
in Ref. \cite{De15} was determined using a Skyrme like energy density functional by imposing the 
constraint on the incompressibility slope parameter at the crossing density ($\sim$ 0.1 fm$^{-3}$) 
determined from energies of the isoscalar giant monopole resonance in the $^{132} {\rm Sn}$ and $^{208} 
{\rm Pb}$ nuclei \cite{Khan2012,Khan2013}}. 

{The above analysis is dependent on the star mass used to
calculate the tidal deformability. However, it is important to
notice that the contribution of $M_0$ to the linear combination
$M_0+\beta L_0$ is maximum for the larger star masses, so 
large star masses that satisfy the $q$ constraints should be
chosen. Taking $\Lambda_{1.6}$ ($q=0.7$) to
constraint $M_0$ the upper limits would have been $\sim 5-10\%$
lower.}

{We have next considered  the range of acceptable values for
$M_0$ just determined,  together with the bounds on
$\Lambda_{1.4}$ and Eq. (\ref{eq_mksym}), to set also constraints on
$K_{{\rm sym},0}$. The results are presented in Table \ref{tab4}: the
ranges $-113<K_{\rm sym,0}< {-52}$ MeV  are obtained for the constraints {on} the symmetry energy
slope from   \cite{Lattimer2013} and $-141 <K_{{\rm sym},0}< {16}$ MeV imposing 
the constraints from \cite{Oertel2017}. The symmetry energy curvature
is a quantity that is still not constrained experimentally. In
\cite{Mondal2017}, the
authors have obtained from the  universality of the correlation structure
between  the different symmetry energy elements and from some well known
nuclear matter properties the range
$K_{{\rm sym},0}=-111.8\pm 71.3$ MeV. Our bounds discussed above are in a
quite good agreement with these values.}
\begin{table}
\label{tab4}
\caption{\label{tab4} The empirical values of $M_0$ and $K_{{\rm sym},0}$ derived for different limits on $\Lambda_{1.4}$ 
         and $L_0$. The bounds on $\Lambda_{1.4}>344$ and $<687(859)$
         obtained from Fig. \ref{fig6_n} are considered. 
         The ranges of $L_0=40-62$ MeV and $L_0=30-86$ MeV are taken from Refs. \cite{Lattimer2013,Oertel2017}.   
         }
\begin{ruledtabular}
\begin{tabular}{lccc}
 $L_0$ & $\Lambda_{1.4}$ & $M_0$ & $K_{{\rm sym},0}$\\
(MeV) &  & (MeV) & (MeV)\\
\hline 
40 – 62  & 344 – 687  &   2254 – 3272   &   -113 – -52     \\
         & 344 – 859  &   2254 – 3631   &   -112 – -52    \\   
30 – 86  & 344 – 687  &   1926 – 3409   &   -141 – 16     \\    
         & 344 – 859  &   1926 – 3768   &   -140 – 16    \\          
\end{tabular}
\end{ruledtabular}
\end{table}    
 
\begin{figure} 
 \centering
\begin{tabular}{c}
    \includegraphics[width=.48\textwidth,angle=0]{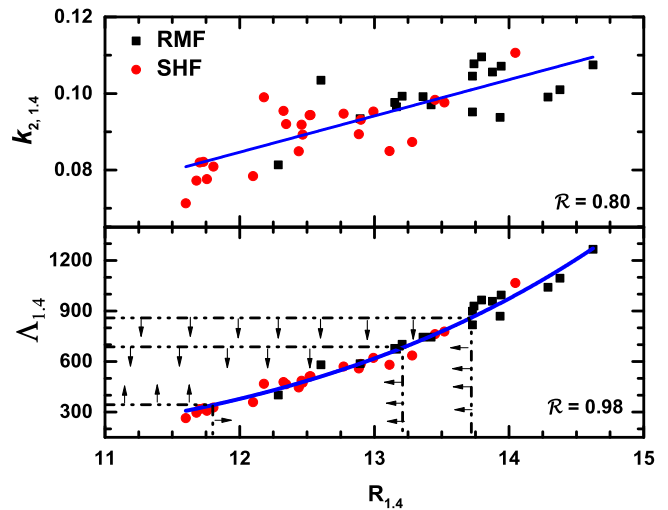}
    \end{tabular}
    \caption{(Color online) {(a) The variation of tidal Love number $k_{2,1.4}$ and (b) the dimensionless tidal 
                            deformability $\Lambda_{1.4}$ with the radius $R_{1.4}$ obtained for 
                            the RMF (black squares) and SHF (red circles) models. The solid lines in the 
                            top and bottom panels are the best fitted
                            linear and curve lines, respectively.{
                            The horizontal dot-dot-dashed lines represent the
                            bounds obtained in Fig. \ref{fig6_n}.}}}

    \label{fig8}
\end{figure}

{Fig. \ref{fig8} displays the tidal Love number $k_{2,1.4}$ (top panel) and the dimensionless tidal 
deformability $\Lambda_{1.4}$ (bottom panel) as a function of NS radius $R_{1.4}$.  
It is evident from the Eq. (\ref{eq2}) that the tidal deformability depends mainly on the 
NS radius and the Love number $k_2$. The $\Lambda_{1.4}$ is expected to be strongly correlated
with $R_{1.4}$ provided either $k_2$ is model independent or it is correlated with $R_{1.4}$.
We observed from Fig. \ref{fig2} that the value of $k_2$ is sensitive to the model 
used which might influence the correlation between $\Lambda_{1.4}$ and $R_{1.4}$. However,
the $k_{2,1.4}$ is moderately correlated with $R_{1.4}$ (top panel) {which ensures 
the persistence of the strong correlation ($\mathcal{R}=0.98$) between $\Lambda_{1.4}$ and $R_{1.4}$ 
(bottom panel).} 
The solid line in the bottom panel represent the fitted curve with equation 
{$\Lambda_{1.4}=9.11 \times 10^{-5} ~ (\frac{R_{1.4}}{\rm
    km})^{6.13}$}. {This equation 
can be rewritten in a form similar to the relation obtained in
\cite{SDe2018}, that expresses the tidal deformability in terms of  the
compactness parameter of the star $\beta= Gm/(Rc^2)$,
$$
\Lambda=a \beta^{-6},
$$
having  the
exponent 6.13 instead of 6. We have verified that the
exponent is mass dependent although close to 6:
taking $m=1.17\, M_\odot$
and $m=1.60\, M_\odot$ the exponent is respectively, {5.84} and
{6.58}. In our analysis we use a set of models different from the one
used in \cite{SDe2018}, and
besides, we have only considered unified inner
crust-core EoS, while in \cite{SDe2018} all the EoS have a common
crust EoS. These two aspects could explain some of the differences.
}Using the derived bounds on $\Lambda_{1.4}$, 
the value of $R_{1.4}$ is found  to be in the range 11.82 – 13.22 (11.82 – 13.72) km for $\Lambda_{1.4}$ 
in the range of 344 – 687 (344 – 859). 
{These ranges for $R_{1.4}$ lie almost within the bounds of 8-14 km and 10.5 - 13.3 km as 
estimated from GW170817 in Refs. \cite{SDe2018,LIGO2}.} 
{Further, our predictions are in harmony with
$R_{1.4} = 11.5$ – $13.6$ km \cite{Li2006}
as constrained by the slope of the symmetry energy which was extracted 
using the terrestrial laboratory data on the isospin diffusion in 
heavy-ion reactions at intermediate energies.}

\section{Conclusions}
            The recent  observation of GW170817 has provided an upper bound on
tidal deformability parameter. {Complementing the gravitation waves observation with the
detection of the  UV/optical/infrared counterpart of GW170817, a
lower bound on  tidal deformability parameter is suggested \cite{Radice2018}.}
We have used a diverse set of relativistic and non relativistic mean field 
models to {look for correlations} of $\Lambda$ with several nuclear matter parameters 
characterizing the {EoS such as the} nuclear matter incompressibility and symmetry energy 
coefficients, and their density derivatives. All the models selected are 
consistent with the bulk properties of finite nuclei as well as with the 
observation of NS with mass of ${\sim 2 \rm M}_\odot$. Nevertheless, across these 
models, the values of $\Lambda$ and {of the} various nuclear matter parameters associated with 
different EoSs vary over a wide range.

      The tidal deformability is found to be weakly or only moderately
correlated with the individual nuclear matter parameters of the EoS. The stronger
correlation of $\Lambda$ is found only for specific choices of the linear
combinations of the isoscalar and isovector EoS parameters. The parameter
$\Lambda$ is  strongly correlated with the linear combination of the
slopes of  incompressibility and symmetry energy coefficients, i.e., $M_0+
\beta L_0$. {Further,} the parameter $\Lambda$ and the Love number
$k_2$ both are strongly correlated with the linear combination of $M_0
+ \eta K_{\rm sym,0}$.
  
      We show that the bound on weighted average of tidal deformability 
for a system of binary neutron star, obtained from complementary analysis 
\cite{GW170817,Radice2018,SDe2018} of GW170817, yields the tidal 
deformability for NS with mass $1.4~{\rm M}_\odot$ in the range of 
$344 < \Lambda_{1.4} < 859$. 
With the aid of the correlations of $\Lambda_{1.4}$ with linear combinations
of nuclear matter parameters as considered together with the bounds on $\Lambda_{1.4}$ 
and the empirical ranges of $L_0$ obtained in Ref. \cite{Oertel2017,Lattimer2013},
we have constrained the values of $M_0$ and $K_{\rm sym,0}$ {to lie in
the intervals $2254<M_0<3631$ MeV and $-112<K_{\rm sym,0}<-52$ MeV or
$1926<M_0<3768$ MeV and $-140<K_{\rm sym,0}<16$ MeV, depending on
the constraints set on $L_0$.} 
The strong correlation of tidal deformability with the NS radius for 
a $1.4 ~ {\rm M}_\odot$ NS yields $R_{1.4}$ in the range
11.82 – 13.72} km. The precise measurement of tidal deformability will provide an
alternative and accurate estimate for $M_0$, $K_{\rm sym,0}$ and $R_{1.4}$. 

\appendix
\section{Mass dependence of $\alpha$, $\beta$ and $\eta$}
\label{apd1}
The coefficients $\alpha$, $\beta$ and $\eta$ are obtained in such a way that they optimize
            the correlations of $\Lambda$, for a given NS mass, with the linear combinations 
            $K_0+ \alpha L_0$, $M_0+\beta L_0$ and $M_0+ \eta K_{{\rm sym},0}$. 
             The value of these coefficients
            are given in Table \ref{tab2} for a few selected NS masses. 
\begin{figure}[htb!]
 \centering
\begin{tabular}{c}
    \includegraphics[width=.45\textwidth,angle=0]{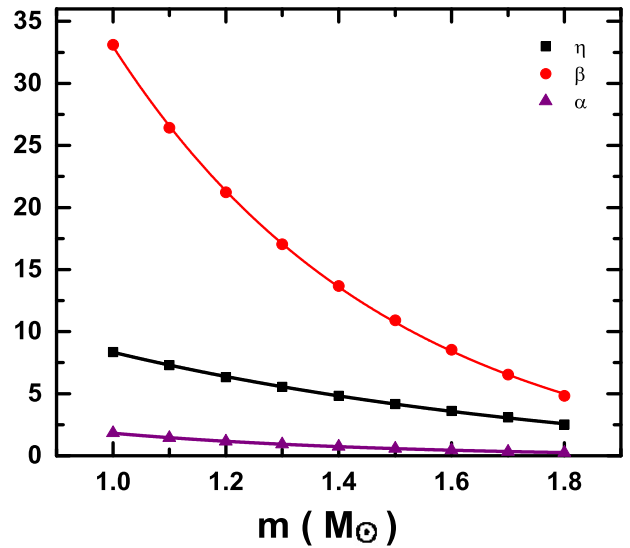}
\end{tabular}
\caption{\label{fig9}(Color online) The values of $\alpha$, $\beta$ and $\eta$ obtained by optimizing the 
                       the correlations of $\Lambda$ with the linear combinations $K_0+ \alpha L_0$,
                       $M_0+\beta L_0$ and $M_0+ \eta K_{{\rm sym},0}$ are plotted as a function of NS mass.}
\end{figure}
            In Figure \ref{fig9}, we plot the mass dependence of $\alpha$, $\beta$ and $\eta$. 
These coefficients can be easily fitted to the 
exponential decay 
like function which can be expressed as $\alpha=-0.13+14.87~ exp(-m/0.49)$, 
$\beta=-1.90 + 265.02~exp(-m/0.49)$ and $\eta= -1.4 +
29.81~ exp(-m/0.89)$, where the NS mass $m$
is in the unit of solar mass. 

\vskip 3in
\begin{acknowledgements}
T.M is grateful to the Theory Group, Saha Institute of Nuclear Physics
for the hospitality accorded to him during the phase of this work. 
The work of M.F. has been partially supported by the NCN (Poland) 
Grant No. 2014/13/B/ST9/02621 and by a STSM grant from the COST 
action CA16214 “PHAROS”. C.P. acknowledges financial support by Fundação para a Ciência e Tecnologia (FCT) Portugal under 
project No. UID/FIS/04564/2016,  project POCI-01-0145-FEDER-029912 with financial support from 
POCI, in its FEDER component, and by the FCT/MCTES budget through national funds (OE), and the 
COST action CA16214 “PHAROS.
T.K.J \& T.M thanks DAE-BRNS for its support (2013/37P/5/BRNS). 
B. K. would like to thank P. Landry and Kenta Hotokezaka for useful 
discussions. N. A. is partially supported by Indo-French CEFIPRA project.
\end{acknowledgements}

%
\end{document}